# Superconductivity at 34.7 K in the iron arsenide $Eu_{0.7}Na_{0.3}Fe_2As_2$

Yanpeng Qi, Zhaoshun Gao, Lei Wang, Dongliang Wang, Xianping Zhang, Yanwei Ma[*]

Key Laboratory of Applied Superconductivity, Institute of Electrical Engineering,

Chinese Academy of Sciences, P. O. Box 2703, Beijing 100190, China

**Abstract:**

$EuFe_2As_2$ is a member of the ternary iron arsenide family. Similar to $BaFe_2As_2$ and $SrFe_2As_2$, $EuFe_2As_2$ exhibits a clear anomaly in resistivity near 200 K. Here we report the discovery of superconductivity in $Eu_{0.7}Na_{0.3}Fe_2As_2$ by partial substitution of the europium site with sodium. $ThCr_2Si_2$ tetragonal structure, as expected for $EuFe_2As_2$, is formed as the main phase for the composition $Eu_{0.7}Na_{0.3}Fe_2As_2$. Resistivity measurement reveals that the transition temperature $T_c$ as high as 34.7 K is observed in this compound. The rate of $T_c$ suppression with the applied magnetic field is 3.87 T / K, giving an extrapolated zero-temperature upper critical field of 90 T. It demonstrates a very encouraging application of the new superconductors.

[*] Author to whom correspondence should be addressed; E-mail: ywma@mail.iee.ac.cn

Realization of high-temperature superconductivity even room-temperature superconductivity is one of the ultimate goals in the field of materials science. The ongoing search for new superconductors has recently yielded a new family of Fe-based compounds $LaFeAsO_{1-x}F_x$ with transition temperatures $T_c$ up to 26 K [1]. By replacing La with other rare earths, $T_c$ can be raised to above 50 K [2-9], and thus the first non-copper-oxide superconductor with $T_c$ exceeding 50 K has emerged. Similar to the cuprates, the Fe-As layer is thought to be responsible for superconductivity and R-O layer is carrier reservoir layer to provide electron carrier. These discoveries have generated much interest for exploring even higher temperature superconductor, and open the new chapter in studies of high-temperature superconductivity outside the well-known domain of copper oxides.

Recently, the iron arsenide $BaFe_2As_2$ in a tetragonal $ThCr_2Si_2$-type structure shows superconductivity at 38 K by hole doping with partial substitution of potassium for barium [10]. The $BaFe_2As_2$ compound is built up with identical Fe-As layers separated by Ba instead of R-O layers [11], so there are double Fe-As layers in unit cell. Then $SrFe_2As_2$ and $CaFe_2As_2$ have been synthesized successfully and superconductivity at ~38 K was discovered by replacing the alkaline earth element with alkali elements [12-15]. $EuFe_2As_2$ is another member of the ternary iron arsenide family [16, 17]. The Eu in $EuFe_2As_2$ is divalent and similar to alkaline earth element in chemical property. $EuFe_2As_2$ exhibits a clear anomaly in resistivity near 200 K [18], which is similar to $BaFe_2As_2$ and $SrFe_2As_2$. It suggests that $EuFe_2As_2$ is another promising parent compound in which superconductivity may be realized by appropriate doping. Very recently, Jeevan et al. reported high-temperature superconductivity in $Eu_{0.5}K_{0.5}Fe_2As_2$, in which K-doping suppresses the SDW transition and in turn gives rise to superconductivity at 32 K [19]. Sodium is similar to potassium in chemical property and $Na^+$ radius is close to $Eu^{2+}$. In this paper, we report the successfully fabrication of $Eu_{0.7}Na_{0.3}Fe_2As_2$ compound by hole doping with partial substitution of sodium for europium. Similar to K-doping, Na-doping strongly weakens the anomaly and induces superconductivity at 34.7 K, which is comparable to the $T_c$ of $Eu_{0.5}K_{0.5}Fe_2As_2$ [19]. The upper critical fields ($H_{c2}$) determined according to the Werthamer-Helfand-Hohenberg formula are (T= 0 K) ≈ 100 T, indicating a very encouraging application of the new superconductors.

**Experimental**

The superconductors with nominal composition of Na-doped $Eu_{0.7}Na_{0.3}Fe_2As_2$ were prepared by one-step solid state reaction. The details of fabrication process are described elsewhere [9]. Stoichiometric amounts of the starting elements Eu (99.99%), Na (99.5%), Fe (99.99%) and As (99.999%) were thoroughly grounded by hand and encased into pure Nb tubes (One end of the tube was sealed). After packing, the other tube end was crumpled, and this tube was subsequently rotary swaged and sealed in a Fe tube. The sealed samples were slowly heated to 850 °C and kept at this temperature for 35 hours. The high purity argon gas was allowed to flow into the furnace during the heat-treatment process. It is note that the sample preparation process except for annealing was performed in glove box in which high pure argon atmosphere is filled.

Phase identification and crystal structure investigation were carried out using x-ray diffraction (XRD) using Cu $K_\alpha$ radiation. Resistivity measurements were performed by the conventional four-point-probe method using a Quantum Design (PPMS). AC magnetic susceptibility of the samples was measured by an Oxford cryogenic system (Maglab-12).

**Results and discussion**

The XRD pattern for the prepared sample is shown in figure 1. It is seen that all main peaks can be indexed by the $ThCr_2Si_2$ tetragonal structure with a = 3.8978 Å, and c = 12.2623 Å. It is noted that the lattice parameter values are a = 3.9104 Å, and c = 12.1362 Å for undoped compound $EuFe_2As_2$ [18]. Clearly Na doping leads to an apparent decrease in a-axis lattice and an increase in c-axis lattice. This result is similar to $Ba_{1-x}K_xFe_2As_2$ [20]. Small amount of FeAs impurity was also observed in the XRD pattern. Such impurity phases might be reduced by optimizing the heating process and stoichiometry ratio of start materials.

Figure 2 shows the temperature dependence of the electrical resistivity for $Eu_{0.7}Na_{0.3}Fe_2As_2$ sample. From this figure we can observe a sharp transition with the onset temperature at 34.7 K. The residual resistivity ratio RRR = ρ (300 K)/ ρ (35 K)

= 4.87, indicating a good quality of our sample. As reported by Ren et al., undoped $EuFe_2As_2$ exhibits a clear anomaly near 200 K [18], which is ascribed to the spin-density-wave instability and structural phase transitions from tetragonal to orthorhombic symmetry. Due to Na-doping, the high-temperature anomaly is suppressed and then superconductivity occurs. It is noted that a slight bump near 200 K is also seen in the resistivity curve. The anomaly is not completely suppressed, which suggests that $T_c$ could be increased further as long as more sodium was doped into $EuFe_2As_2$.

In order to further confirm the superconductivity of $Eu_{0.7}Na_{0.3}Fe_2As_2$, AC magnetic susceptibility measurement also was performed. Figure 3 shows the temperature dependence of AC magnetization with the measuring frequency of 333 Hz and the amplitude of 0.1 Oe. The sample shows a well diamagnetic signal and superconductivity with $T_c$ =32 K, which is corresponding to the middle transition point of resistance. The sharp magnetic transitions on AC curves indicate the good quality of our superconducting samples. Estimation on the magnetic signal indicates that the superconducting shielding volume of the sample is beyond 90%.

We carried out the resistance versus temperature measurement for the $Eu_{0.7}Na_{0.3}Fe_2As_2$ sample under different magnetic fields. The magnetic field is observed to suppress the transitions as expected for a superconducting transition, the onset transition point and zero resistance point shift to lower temperatures. We tried to estimate the upper critical field ($H_{c2}$) and irreversibility field ($H_{irr}$), using the 90% and 10% points on the resistive transition curves. Figure 4 shows the temperature dependence of $H_{c2}$ and $H_{irr}$ with magnetic fields up to 9 T for the $Eu_{0.7}Na_{0.3}Fe_2As_2$ sample. It is clear that the curve of $H_{c2}$ (T) is very steep with a slope of $- dH_{c2}/dT|_{Tc}$ = 3.87 T / K. From this Figure, using the Werthamer-Helfand-Hohenberg formula [21], $H_{c2}(0) = 0.693 \times (dH_{c2}/dT) \times T_c$. Taking $T_c$= 33.7 K, we can get $H_{c2}(0) \approx 90$ T. If adopting a criterion of 99 %$\rho_n$(T) instead of 90%$\rho_n$(T), the $H_{c2}(0)$ value of this sample obtained by this equation is higher than 100 T. We can see that the irreversibility field is rather high compared to that in $MgB_2$. These high values of $H_{c2}$ and $H_{irr}$ indicate that this new superconductor has an encouraging application in very high fields.

According to the relationship between $H_{c2}$ and the coherence length ξ, namely, $H_{c2}= \Phi_0/(2\pi\xi^2)$, where $\Phi_0$ is the flux quantum, the value of the coherence length is estimated to be ~19 Å ($H_{c2}(0) \approx 90$ T).

As we know, $ThCr_2Si_2$-type superconductor, $AFe_2As_2$, is another family member of Fe-As superconductors. By replacing the alkaline earth elements with alkali elements, superconductivity was discovered in $BaFe_2As_2$, $SrFe_2As_2$ and $CaFe_2As_2$ [10-15], which caused an excitement in the scientific community. The structure of $AFe_2As_2$ superconductor is much simpler compared to that of ZrCuSiAs-type ones. More excitingly, it is much easier to synthesize because of without oxygen, and large single crystals could be obtained by self-flux method, which is important to investigate the intrinsic property of Fe-As superconductor. However, all *A* elements come from alkaline earth elements. Very recently, Jeevan et al. discovered high-temperature superconductivity in $EuFe_2As_2$ [18] and now we report the superconductivity of $Eu_{0.7}Na_{0.3}Fe_2As_2$ compound. It clearly demonstrates that $EuFe_2As_2$ is a new member of $ThCr_2Si_2$-type Fe-As superconductor [22], which offers more chances to study the mechanism of new iron-based superconductor.

**Conclusions**

To summarize, we have successfully synthesized the iron-based Na-doped layered compound $Eu_{0.7}Na_{0.3}Fe_2As_2$ by one-step solid state reaction method. XRD diffraction shows that $Eu_{0.7}Na_{0.3}Fe_2As_2$ has $ThCr_2Si_2$ tetragonal structure with a = 3.8978 Å, and c = 12.2623 Å. Na-doping leads to suppression of the anomaly partially in resistivity and induces superconductivity at 34.7 K. Furthermore, the upper critical fields ($H_{c2}$) determined according to the Werthamer-Helfand-Hohenberg formula are (T= 0 K) ≈ 100 T, indicating a very encouraging application of the new superconductors.

The authors are grateful to Profs. Haihu Wen, Liye Xiao and Liangzhen Lin for stimulating discussion and useful advices. This work was partly supported by the Natural Science Foundation of China (Contract Nos. 50572104 and 50777062) and National '973' Program (Grant No. 2006CB601004).

**Captions**

Figure 1 XRD patterns of the $Eu_{0.7}Na_{0.3}Fe_2As_2$ sample. The impurity phases are marked by *.

Figure 2 Temperature dependences of resistivity for the $Eu_{0.7}Na_{0.3}Fe_2As_2$ sample measured in zero field. Inset: Enlarged view of low temperature, showing superconducting transition.

Figure 3 Temperature dependence of magnetic susceptibility measured with $H_{ac}$ = 0.1 Oe, f = 333Hz.

Figure 4 The upper critical field $H_{c2}$ and $H_{irr}$ as a function of temperature for $Eu_{0.7}Na_{0.3}Fe_2As_2$ samples. The $H_{c2}$ and $H_{irr}$ values were defined as the 90% and 10% points of the resistive transition, respectively.

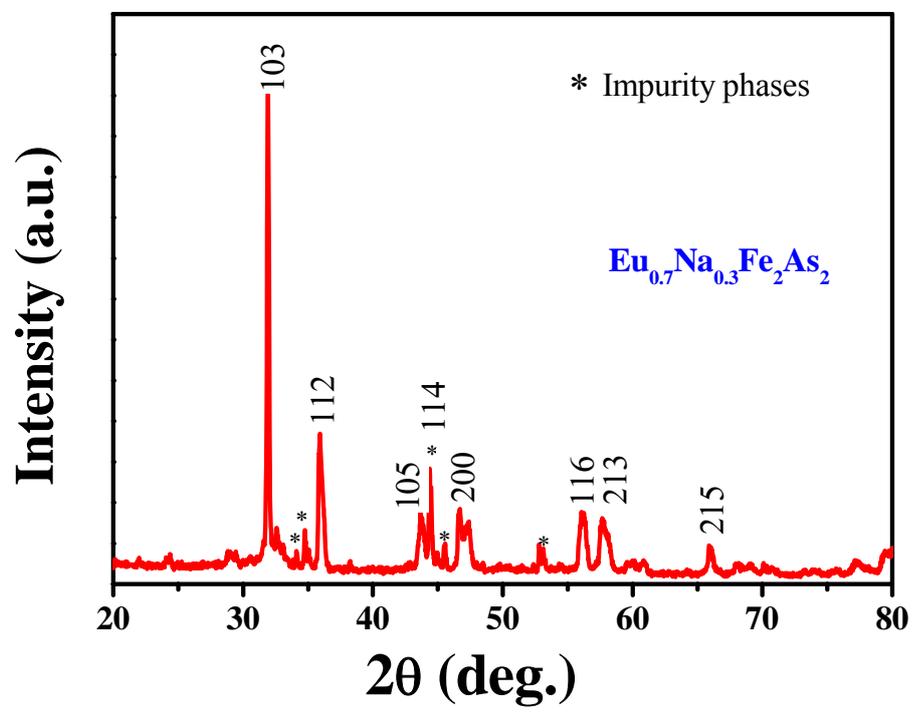

Fig.1 Qi et al.

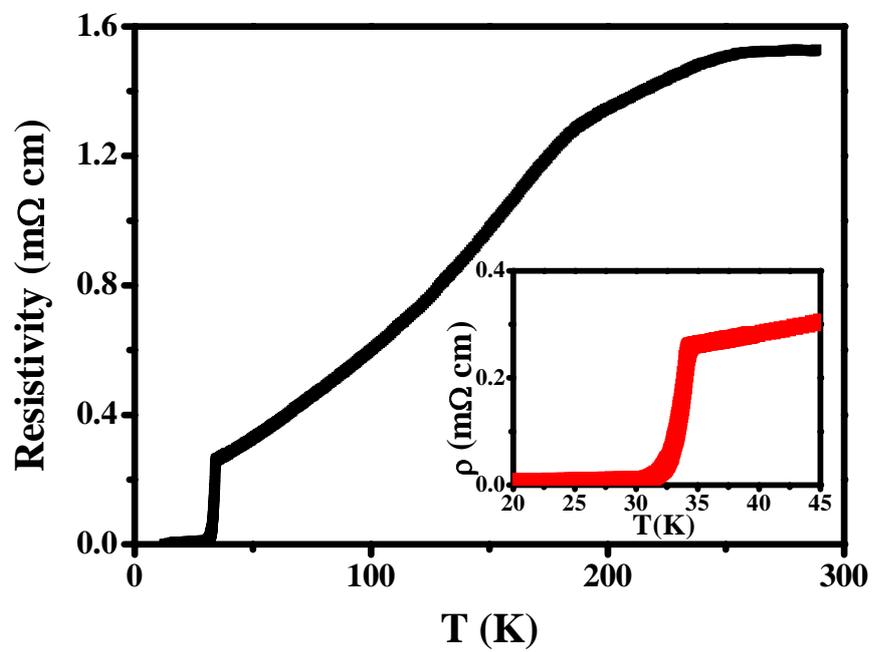

Fig.2 Qi et al.

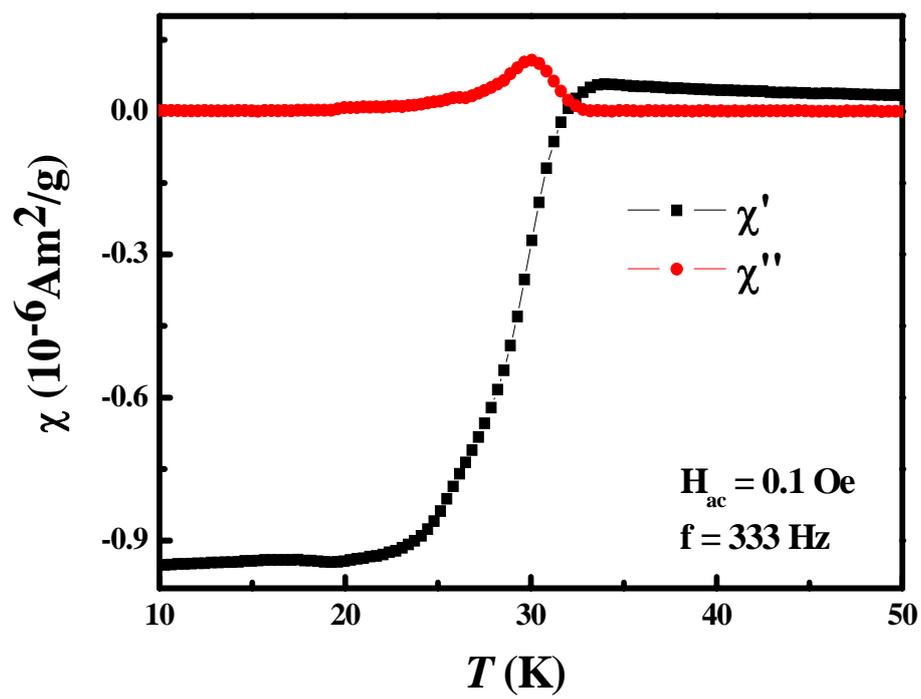

Fig.3 Qi et al.

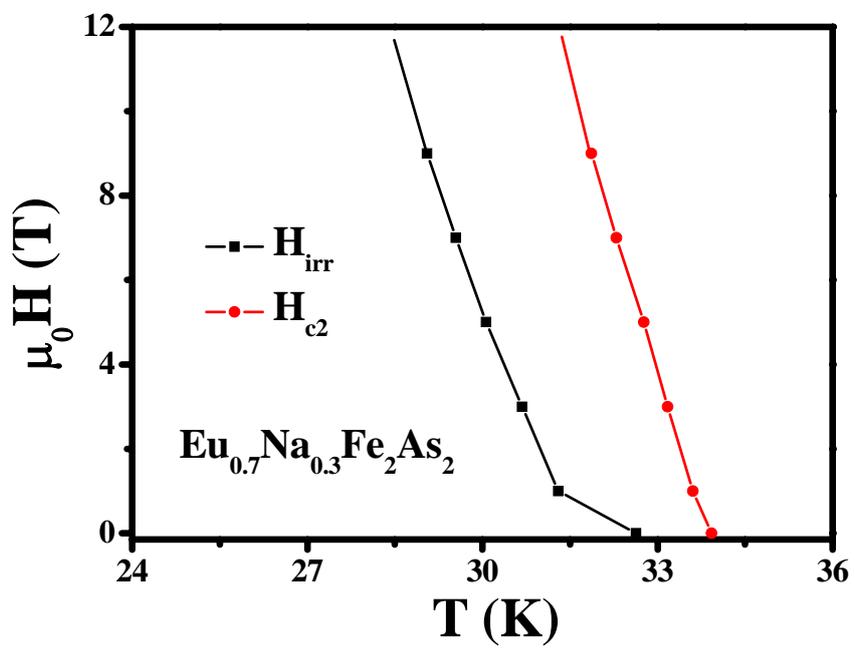

Fig.4 Qi et al.